\documentclass[sigconf]{acmart}

\usepackage{framed}
\usepackage[strict]{changepage}
\definecolor{formalshade}{rgb}{0.95,0.95,1}
\definecolor{darkblue}{rgb}{0.36,0.54,0.66}

\newenvironment{formal}{%
  \MakeFramed{\advance\hsize-\width\FrameRestore}%
  \noindent\hspace{-4.55pt}
  \begin{adjustwidth}{}{7pt}%
  \vspace{2pt}\vspace{2pt}%
}
{%
  \vspace{2pt}\end{adjustwidth}\endMakeFramed%
}

\AtBeginDocument{%
  \providecommand\BibTeX{{%
    \normalfont B\kern-0.5em{\scshape i\kern-0.25em b}\kern-0.8em\TeX}}}

\setcopyright{ccbysa}
\copyrightyear{2023}
\acmYear{2023}
\acmDOI{}

\acmConference[Generative AI and HCI '23 Workshop]{Generative AI and HCI}{April 28,
  2023}{Hamburg, Germany}
\acmBooktitle{Generative AI and HCI '23 Workshop, April 28, 2023, Hamburg, Germany}
\acmISBN{}

\begin{document}

\title[Rolling the Dice]{Rolling the Dice: Imagining Generative AI as a \\ Dungeons \& Dragons Storytelling Companion}

\settopmatter{authorsperrow=4}
\author{Jose Ma. {Santiago} III}
\orcid{0002-4730-7865}
\affiliation{%
  \institution{Paris-Lodron Universität Salzburg}
  \city{Salzburg}
  \country{Austria}}
\affiliation{%
  \institution{Salzburg University of Applied Sciences}
  \city{Salzburg}
  \country{Austria}}
 \email{jose.santiago@stud.sbg.ac.at}

 \author{Richard Lance {Parayno}}
\orcid{1234-5678-9012}
\affiliation{%
  \institution{Paris-Lodron Universität Salzburg}
  \city{Salzburg}
  \country{Austria}}
\affiliation{%
  \institution{Salzburg University of Applied Sciences}
  \city{Salzburg}
  \country{Austria}}
 \email{richard.parayno@stud.sbg.ac.at}

\author{Jordan Aiko {Deja}}
\orcid{0001-9341-6088}
 \affiliation{%
  \institution{De La Salle University}
  \city{Manila}
  \country{Philippines}}
\affiliation{%
  \institution{University of Primorska}
  \city{Koper}
  \country{Slovenia}
 \postcode{6000}}
\email{jordan.deja@dlsu.edu.ph}

\author{Briane Paul V. {Samson}}
\orcid{0002-0253-452X} 
 \affiliation{%
  \institution{De La Salle University}
  \city{Manila}
  \country{Philippines}}
\email{briane.samson@dlsu.edu.ph}


\begin{abstract}
AI Advancements have augmented casual writing and story generation, but their usage poses challenges in collaborative storytelling. In role-playing games such as Dungeons \& Dragons (D\&D), composing prompts using generative AI requires a technical understanding to generate ideal results, which is difficult for novices. Thus, emergent narratives organically developed based on player actions and decisions have yet to be fully utilized. This paper envisions the use of generative AI in transforming storytelling into an interactive drama using dynamic and immersive narratives. First, we describe scenarios where narratives are created and character conversations are designed within an overarching fantasy disposition. Then, we recommend design guidelines to help create tools using generative AI in interactive storytelling. Lastly, we raise questions on its potential impact on player immersion and cognitive load. Our contributions may be expanded within the broader interactive storytelling domain, such as speech-conversational AI and persona-driven chatbots.
\end{abstract}





\begin{teaserfigure}
\centering
  \includegraphics[width=1\columnwidth]{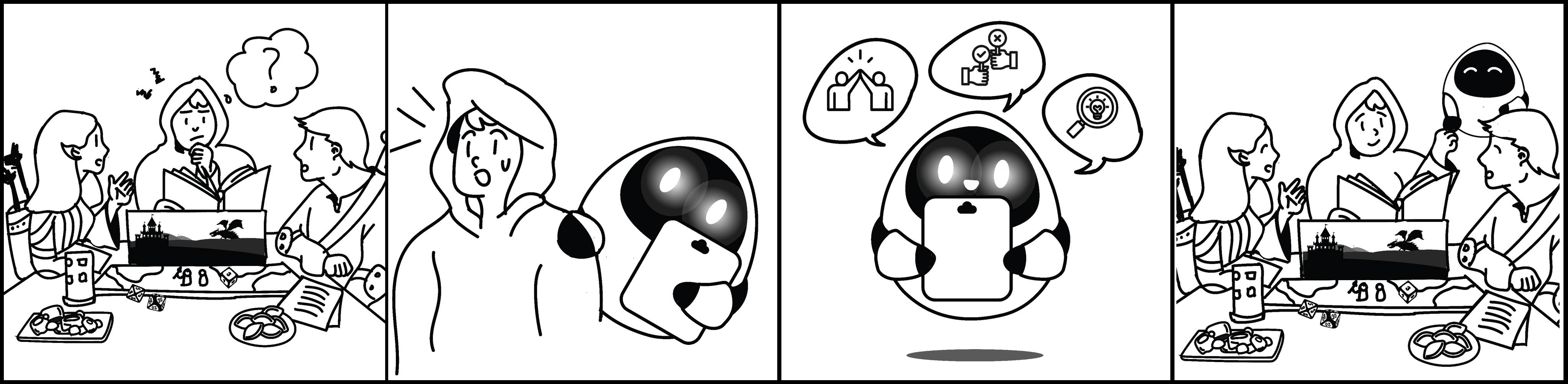}
  \caption{\textbf{A future where AI works along side dungeon masters as a storytelling companion in Dungeons \& Dragons.}}
  \Description{lol}
  \label{fig:teaser}
\end{teaserfigure}
\maketitle
\section{Introduction and Background}

\par In recent years, AI-based collaborative writing tools have been rolled out in various application areas such as argumentative writing~\cite{lee_coauthor_2022}, code synthesis~\cite{austin_program_2021}, drama~\cite{flowers_gamemasters_2023} and storytelling~\cite{fotedar_storytelling_2020, yuan_wordcraft_2022}. These systems are built using large Language Models (LMs) that take prompts from the user and return plausible continuations~\cite{brown_language_2020}. With the use of these tools, authors who struggle most especially during a ``writer's block'' may draw inspiration and benefit from the help of these tools. The writer takes control of their craft by deciding on their persona, style and identity in their written work. Most especially in drama and storytelling, conveying a narrative through spoken or written word requires the presence of culture and tradition~\cite{greene_storytelling_1996}, which is usually not seen in most AI-generated content.

\par Specifically, in the context of tabletop role-playing games (TTRPGs) like Pathfinder and Dungeons \& Dragons (D\&D), a dungeon master (DM) directs the flow of a game's story and leaves room for the players to interact using personas that fit the context of the narrative~\cite{walden_living_2015}. This is done throughout a campaign divided into game sessions, usually held weekly. This is different from traditional storytelling, where the practice of conveying a narrative or message through spoken or written words is a critical component~\cite{greene_storytelling_1996}. 


\par While this setup has been the standard practice more recently, despite the platform undergoing multiple iterations or editions~\cite{walden_living_2015}, it is unclear whether this is an ideal or usable approach. In a study by Reinikainen~\cite{reinikainen2008role}, players encounter difficulty immersing in the game due to the complexity of the recent editions. Such factors may affect the overall experience of the game, especially its usability. The players pointed out that the lack of immersion led to diminished enjoyment and increased frustrations, which undermined the reason why they played. This leaves more room for improvement towards better immersive experiences for players and narrators while reducing cognitive load. 
\par Recent work in generative AI has tried to address these gaps through collaborative storytelling, from using LMs to spark creativity in authors~\cite{yuan_wordcraft_2022, osone_buncho_2021} to generative games~\cite{kreminski_cozy_2019}. This is further improved by recent developments in Natural Language Processing (NLP). Storytelling AI aims to generate engaging narratives with visual illustrations to immerse the user with minimal user input \cite{fotedar_storytelling_2020}. The tool creates narrative text and complements images by combining two generative models. By using new large LMs such as GPT-3, we also have the added functionality of text-to-image generation \cite{ramesh_zero-shot_2021}. With the two components set, we can tackle the question, \textit{how can generative AI enhance the creativity and flexibility of D\&D storytelling?} Thus, this paper presents the following contributions: (1) We provide sample scenarios in D\&D where we use generative AI (using prompts) as a form of collaborative storytelling towards improved user engagement. (2) We provide design guidelines towards expanding the space of immersive collaborative AI-based storytelling scenarios in D\&D and beyond. 

\section{D\&D Scenarios}
\par With the possibilities afforded by large LMs, we imagine storytelling scenarios in D\&D with an AI companion named \textit{Avalon}. We provide sample prompts and outputs and discuss them in the form of guidelines in the succeeding sections. 

\subsection{Scenario 1: Where it begins...}
\begin{formal}
    Prompt 1: The adventurers reach a fork in the dungeon, but instead of picking one of two paths, they make a third path"
\end{formal}
\par As a dungeon master, you run through the dungeon that will be the initial setting for your first D\&D session, as it is your job to control the flow of the narrative of the session~\cite{walden_living_2015}. As your players eventually arrive, you go through your notes one last time as they settle down and get into character. You elaborately describe the scenario they are in \textit{waking up in a dimly lit dungeon with two paths before them.} You can see from their faces the shock from your revelation, as one of them proposes they should get out as soon as possible. You ask them what they wish to do, and Gunter the dwarf replies: I use my bombs to make a third path! You did not prepare for this, but your observant companion did. You quickly scramble to describe to Avalon the current situation (see first prompt in quote). Your companion has been intently listening to the actions of the players and was able to incorporate the actions of Gunter the dwarf and how it affects the world around him to match. This way, their story progresses while incorporating narrative techniques to keep the players engaged as they dredge forward into the dungeon!

\subsection{Scenario 2: Continuing where you last left...}
\begin{formal}
    Prompt 2: ``Can you give me merchants with their own personalities from the merchant capital in my D\&D campaign?''
\end{formal}
\par After the party had defeated the gorgon, the keeper of the treasure of Lord Nuk'thali, they decided to visit the nearest city to purchase new gear. You have had the idea of a merchant capital, bustling with trade and commerce, where merchants and adventurers are free to deal as they please. You know that your players would like to chat with the locals and barter for goods, but you need help coming up with names and personalities for each merchant. You rack your brain, trying to think of familiar-sounding names of humans that you never considered that a merchant capital would likely have a more diverse group of locals. You look over to your companion as you describe your vision of the city to them (see second prompt in quote). You are amazed by the diversity of the cast that your companion has provided. It has sparked creativity within you to start writing their backstories and some dialogue based on their personalities. And with that, you are prepared for the next session, eager to find out what your players think of the merchants that await them!

\subsection{Scenario 3: Here we go again...}

\begin{formal}
    Prompt 3: ``Can you generate the art for the fierce battle between the adventuring party and Lord Nuk'thali in the crypt of bones, fantasy style..''
\end{formal}
\begin{figure*}[t]
  \centering
  \includegraphics[width=1\linewidth]{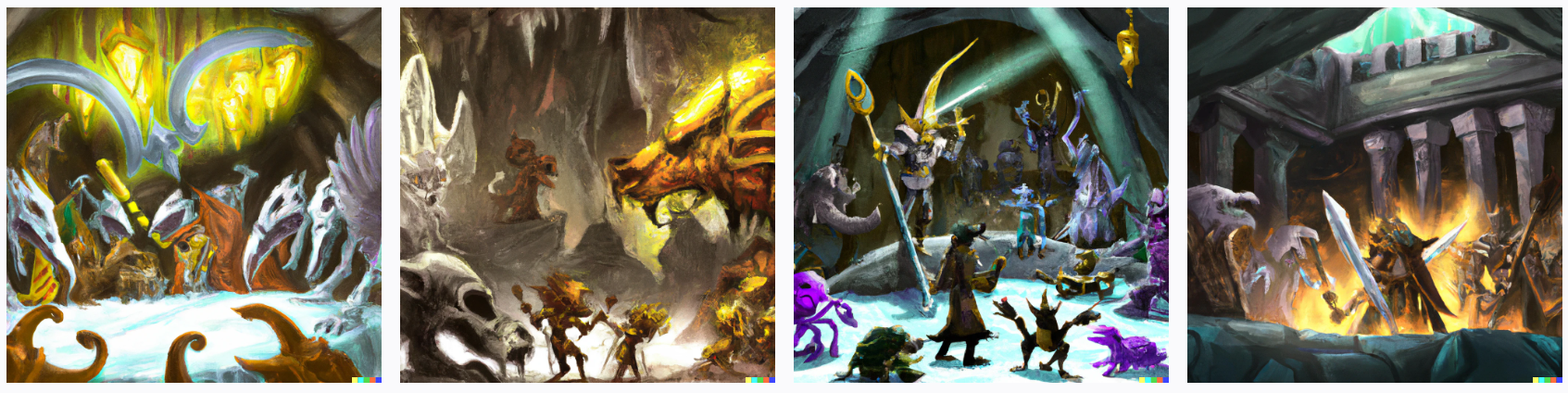}
  \caption{Generated images based on the bottom prompt shown in the quotes. This image was generated using DALL-E (\url{https://openai.com/dall-e-2/}). }
  \Description{Insert image of Dall-e creation}
  \label{fig:dalle}
\end{figure*}

\par It is the last session of the campaign, as you prepare the scene of the final battle with Lord Nuk'Thali. To the eyes of a mortal, it is but a crudely made landscape with cardboard spikes. But to the eyes of the players, it is the Crypt of Bones, lair of the evil lord itself. Eventually, your players arrive marvelling at the scenery as the culmination of all of their efforts thus far. They get into character as you describe what lies before them. Slowly, a figure steps forward, revealing itself to be none other than Lord Nuk'Thali, the final boss and master of the crypt. In the battle, the adventurers face the powerful lich Lord Nuk'Thali and endure a fierce exchange of attacks that lasts for a while. However, they ultimately manage to destroy his phylactery, which leads to his defeat and disintegration. The adventurers emerge victorious but exhausted from the battle. You peek over your DM screen as you see the faces of your players as they cheer and kick after the the most exciting battle they ever had. You know they will remember this campaign forever. To commemorate the event, you look over your companion and asked the third prompt. Given that LMs are capable of remembering previous prompts and answers \cite{brown_language_2020}, your companion was able to generate a scenery (~\autoref{fig:dalle}) that incorporates the story so far. Your party gathers to inspect the illustration as they all decide to frame it, as a remembrance of their very first Dungeons \& Dragons campaign. Eventually, you start a new campaign with a new group accompanied by their own challenges. Despite having run multiple sessions with various groups of players, you can still look over your dining room table and reminisce the experiences you had looking at the framed portrait on the wall.


\section{Discussion}
\par Like other AI collaborative writing tools, AI wears different hats at each stage of the writing process~\cite{yuan_wordcraft_2022} such as an idea generator, scene interpolator and copy editor. In D\&D, however, the narrative is built on the spot as the players explore the fantasy world. Thus, some hats require more attention, precisely that of an idea generator and scene interpolator. From the scenarios above, it is not feasible for the DM to prepare for everything that can happen during a D\&D session. This is why we turn to tools and companions to aid the common shortcomings and pitfalls they encounter. By giving the role to the AI as an idea generator and scene interpolator, we avoid scenarios where they get stuck on ideas and narratives. We present some guidelines to prepare developers and designers to help casual players and writers in similar contexts. 

\subsection{Understanding Context}
\par D\&D vastly differs from generative AI platforms such as ChatGPT~\cite{nakano_webgpt_2022}. Thus, there is a need to have a good understanding of the mode of communication used in TTRPGs. This can be done by analyzing campaigns such as Critical Role, an unscripted live stream of voice actors playing D\&D led by DM Matthew Mercer~\cite{critrole}. The Critical Role Dungeons \& Dragons Dataset (CRD3) analyzed 159 episodes across two campaigns that resulted in 398,682 turns of dialogue~\cite{rameshkumar_storytelling_2020}. By understanding this mode of communication, designers can empathize with the users they are designing for. With the current design of LMs, we are limited to $1:1$ interactions with the user. Incorporating multiple users in a collaborative AI environment may create responses that align with the users' expectations. Audio should also be considered a potential input method, as conversations between players and even with the DM could be crucial to understanding the context. However, further improvements with the structure of LMs might be needed to filter out unnecessary inputs or prompts as chatter is quite frequent in D\&D sessions as seen in the dataset~\cite{rameshkumar_storytelling_2020}.
\par However, understanding the context of the narrative continues beyond just the AI~\cite{sidhu_pivotal_2021}. Another aspect that should be considered is how to help the user understand the context. This can be done in one of two ways, either directly or indirectly. The first method can be done by the narrator, such as describing the scenery of the landscape the players are exploring. However, the narrative becomes stale and boring if the players are just given the information. This is where the second method can be utilized, which can be done through NPCs and other entities in the world. The players could further understand the world from a local perspective by conversing with them. Posters, journals, and even gossip can indirectly give the players more context. This guideline tackles the issue of understanding the context and the mode of communication. Since D\&D uses a setup that involves multiple users talking at the same time, an understanding of the mode of communication is important to grasp the context of the game. With this, we can increase the chances of generating the desired output when consulting with an AI collaborator. 

\subsection{Putting Value into Inspirations}
\par It is common practice that DMs turn to existing forms of traditional storytelling, such as fantasy novels and movies, for inspiration~\cite{slavicsek_dungeon_2006}. DMs often incorporate existing themes, plot hooks, villain motivations, and even character tropes in the context of their fantasy world. By utilizing LMs to integrate themes from an extensive collection of fantasy novels and other forms of media, we can reduce predictable narratives from occurring frequently. While this does not eliminate these instances, having a diverse collection of references proves useful for many reasons. Diversity and inclusion in D\&D are other aspects rarely touched upon but have significantly changed in recent years. A comparative study illustrates how official D\&D content (pre-made adventures, source books, etc.) has struggled with the presentation of minority groups with each iteration~\cite{long_2016_character}. This can be addressed by having a diverse repertoire of literary works. Recently, there has been a challenge to incorporate diversity in science fiction and fantasy-based young adult literature~\cite{garcia_worlds_2017}. Diversity impacts not only the nature of the narratives that can be generated but also the overall experience of the players and DMs. 
\par Another dimension that is worth exploring further is the presence of organic inspiration. DMs have often used real-life experiences to create diverse narratives that incorporate their tastes. This is especially important when envisioning the fantasy world's NPCs and their dialogue and personalities. Thus, generative AI must also be able to consider these attributes in their implementation. Existing LMs can empathize and respond through affective text~\cite{ghosh_affect-lm_2017}. However, emotion encoding is not yet well defined, and some studies have linked our emotional response to our personality types~\cite{gilboa_personality_1994}. Several models on these can be explored further to consider these dimensions. This guideline highlights the importance of inspiration from a diverse collection of works. We aim to limit the occurrence of stale narratives that have been exhausted in previous works and encourage the exploration of emotion encoding and how it can be refined in generative AI.
\subsection{Maximizing Engagement}
\par Meaningful play is the idea that games can create experiences beyond the typical leisurely activity~\cite{tekinbas_rules_2003}. Relationships between player actions and the outcome must be discernible and integrated into the larger context of the game. Player agency is a concept in game studies as the fundamental capability of the player to act in the world~\cite{stang_this_2019}. This concept also applies in D\&D as players and DMs shape the narrative. Meaningful play and player agency work together to engage and immerse the players in the world. One approach that can be used is by emphasizing the interactive world. In D\&D, the theatre of the mind is a concept where players or the DM would describe a scene in vivid detail using the participants' imagination to visualize the scenario they are in~\cite{svan_emergent_2021}. If the AI can describe the scenario the players are in and hint about key areas that players can explore or interact with, then their curiosity would draw them in. It helps by adding pauses or asking the players what they want to do in that situation. This ensures that player agency is still present despite the AI controlling the flow of the narrative. 
\par Another approach would be to highlight the consequences of each player's action in the world. This can range from small cause-and-effect narratives, such as a player cutting down a tree in a forest, causing a loud noise and wildlife to flee from it to more lasting ones such as the death of a character. While one can argue that death is superficial in the context of a fantasy narrative, the actions and experiences that lead to death add weight to the player's choices~\cite{flynn-jones_dont_2015}. A pre-play interview was conducted to understand meaningful play in D\&D and reports that death was the most common meaningful play experience~\cite{sidhu_pivotal_2021}. This aspect in D\&D can potentially transcend the traditional definition of meaningful play to deeper emotions, thus affecting engagement. 
\par We expand further by incorporating external aids. In one of the scenarios, we portrayed the final session of a D\&D campaign, where players eventually defeat the villain of the narrative and save the day. This is usually portrayed in an epic battle of conflicting morals, where the players' actions coalesce into a magnificent display in the theatre of the mind. Using this battle, the companion in the scenario was able to generate illustrations that captured the essence and emotions of players during the fight. This adds engagement outside the narrative by using visual aid such as illustrations, audio and even tactile aid to immerse the players in the narrative. By incorporating these approaches to the narrative style created by the AI and the DM, we expect players to feel more connected to the overarching fantasy world. These aspects can be visualized as threads: the more threads the player has (player agency, understanding of consequences, emotional ties), the stronger the player's connection with the world and the narrative. By ensuring that player agency is present and has reasonable consequence, we can further immerse the players within the setting of the overarching fantasy world. 

\section{Conclusion}
\par As we move towards innovations in the generative AI landscape, we urge designers and developers to reflect on the users rather than the capabilities of the tools. We have the technology needed to create engaging experiences based on existing research. Yet, more work must be done before we deploy them to our users. In this opinion piece, we used the example of Dungeons \& Dragons as a domain where non-technical users may reside. By envisioning the use of generative AI, we push the boundaries by designing systems that enable our users to utilise unfamiliar tools that may be able to help them. The scenarios follow you as the dungeon master as you prepare and conduct D\&D sessions with your playmates. We highlighted common pain points that DMs encounter and described how generative AI could help. From these scenarios, we created an initial list of design guidelines (e.g. understanding context, putting value into inspirations, maximising engagement) that may help developers and designers integrate generative AI for interactive storytelling purposes. Our design guidelines focus on the interaction between AI, the DM and the other players in the game. We call for future explorations on implementing and validating whether the design guidelines create better immersive experiences in broader application areas outside of D\&D. We also look at how this aids in designing collaborative storytelling as a tool for non-technical writers to utilise in their works. As designers, we have the task of evaluation and empathy, regardless of the application area. We hope that with the perspective of D\&D dungeon masters and players, we can work towards a user-centered future in generative AI.
\bibliographystyle{ACM-Reference-Format}
\bibliography{sample-base}


\end{document}